# Flexibility Evaluation of Domestic Electric Water Heater Aggregates

F. Conte, B. Gabriele, S. Massucco, F. Silvestro
DITEN – Università degli studi di Genova
Genova, Italy
fr.conte@unige.it

D. Cirio, L. Croci
Ricerca sul Sistema Energetico – RSE S.p.A.
Milano, Italy
diego.cirio@rse-web.it

*Abstract*—In this paper, a method to evaluate the flexibility of aggregates of domestic electric water heaters is proposed and applied to the Italian case. Flexibility is defined as the capability of the aggregate to vary its power demand for a given time interval. The evaluation method consists of a Monte Carlo analysis, that uses the thermal model of electric water heaters and a proper elaboration of the external inputs, such as ambient and cold water temperatures, and hot water demand. The case of large aggregates defined along the Italian territory has been studied showing the dependence of flexibility on seasons and on time.

## I. Introduction

One of the solutions studied by academic and industrial communities to allow the definitive integration of renewable energy sources (RES) into the electrical power system is the so called demand response (DR). The idea is to ask electrical loads to provide regulation services by modulating their energy consumption, in order to support the power system operation.

Because of their nature, loads have the principal issue of satisfying the end users' demand. Therefore, a fundamental requirement for a load to deliver any regulation service is to be provided with any form of energy reserve, that can be used to vary its power consumption without compromising the end use specifications. In other words, they have to be *flexible*. Loads related to temperature control, such as air cooling and heating systems, fridges, and electric water heaters (EWHs) are clear examples of flexible loads, since they keep a thermal energy reserve that can be exploited to provide regulation services.

Anyway, even if a load is flexible, to realize a DR service is not straightforward. Indeed, differently form large traditional generators, that today sustain the power system operation, loads are higher in number, smaller (in terms of power) and extremely distributed. Moreover, they are usually hard to be monitored and controlled. To cope with these problems, loads have to be gathered into *aggregates*. An aggregate is set of distributed generation and/or load units that cooperate and interface as a unique entity with services and energy markets. Literature provides many examples of load aggregate management strategies with the aim of providing regulation services [1]-[6].

Following [7] and [8], where the European Commission defines the guidelines to allow aggregates to be involved in the dispatching and balancing markets, deliberation [9] of the Italian Regulatory Authority for Energy, Networks and Environment, has led the Italian Transmission System Operator (TSO) to write the codes of pilot projects for the "enlargement of the dispatching resources". In these projects, aggregated entities are called Virtual Qualified Units (VQUs), which can be composed by loads [10], generating units, or by a mix of them [11]. For the latter, fifteen perimeters of aggregation, defined as sets of districts, have been defined.

The dispatching and balancing services designed in [11], require a variation from a baseline profile of the power generation or demand of the VQU for a given time extension, from a minimum of 15 minutes up to 3 hours. The objective of the present work is to evaluate the capability of domestic EWHs to contribute to the provision of such services, as a part of a VQU.

A method to evaluate an EWH aggregate flexibility, based on a Monte Carlo simulation analysis, is introduced in Section II. This method takes into account the effects on flexibility of external inputs such as ambient and cold water temperatures, and the users' hot water demand. In Section III, the Italian case study is detailed. Results are provided in Section IV and commented in Section V, by focusing on the dependence of flexibility on seasonal changes and on time.

## II. Flexibility Evaluation Method

In this study, the flexibility of an aggregate of electrical loads is defined as the capability to temporarily vary its power consumption with respect to a baseline profile, in order to provide a regulation service required by the TSO. It can be positive, meaning that the power consumption is increased, or negative, meaning that the power consumption is reduced.

Flexibility can be therefore determined by the quantities $\Delta P_A^+(\Delta t, \tau)$ and $\Delta P_A^-(\Delta t, \tau)$, defined as the *maximum positive*

This work has been financed by the Research Fund for the Italian Electrical System in compliance with the Decree of Minister of Economic Development April 16, 2018.

and negative power consumption variations achievable starting at time $\tau$ for the time interval $\Delta t$.

The objective of this paper is to define a method to evaluate $\Delta P_A^+(\Delta t, \tau)$ and $\Delta P_A^-(\Delta t, \tau)$ in the case of aggregates of EWHs.

*A. EWHs aggregate simulation*

The flexibility evaluation method proposed in this paper is based on the Monte Carlo simulation of the thermal dynamics of aggregates of EWHs. The adopted model is the following [12]:

$$\dot{T} = -\frac{1}{RS_wV\rho}(T - T_e) - \frac{w}{60V}(T - T_o) + \frac{qP^n}{S_wV\rho}, \quad (1)$$

where: $T$ is the hot water temperature [°C]; $T_e$ is the external ambient temperature [°C]; $T_o$ is the incoming cold water temperature [°C]; $w$ is the hot water withdrawal [l min$^{-1}$]; $R$ is the thermal resistance of the boiler shell [°C/W]; $S_w$ is the water specific heat capacity [J kg$^{-1}$ °C$^{-1}$]; $V$ is the boiler capacity [l]; $\rho$ is the water density [kg l$^{-1}$]; $P^n$ is the EWH nominal power [W]; $q$ is the state of the thermostat (1 = on, 0 = off). This last follows the rule

$$q = \begin{cases} 0 & \text{if } T > T_{sp} + \Delta \\ 1 & \text{if } T < T_{sp} - \Delta \\ q & \text{otherwise} \end{cases}, \quad (2)$$

where $T_{sp}$ is the hot water temperature set-point [°C] and $2\Delta$ is thermostat dead-band [°C].

Let us consider an aggregate of EWH composed by $M$ sub-aggregates, each one characterized by ($i = 1,2,...,M$):
a) the same class of EWHs, defined parameters $V_i$, $R_i$, $P_i^n$;
b) the sub-aggregate nominal power $P_{sA,i}^n$ [W];
c) the average daily profile of the outside ambient temperature $\{\overline{T}_{a,i}\}$ and of the cold water temperature $\{\overline{T}_{o,i}\}$.

It is moreover assumed that all EWHs belonging to the aggregate have:

d) the same average daily profile of the hot water withdrawals $\{\overline{w}\}$;
e) the same maximal temperature limit $T^{max}$;
f) the same thermostat dead-band $2\Delta$;
g) a fixed desired temperature $T^*$ required by users (the standard value adopted in this paper is $T^* = 40 \, °C$).

Given these data, the daily average power consumption profile of the $i$-th sub-aggregate is computed by simulating, with a given time granularity $t_s$ [min] (1 min in this paper), a set of $N$ sample EWHs, by generating $N$ sample profiles of the daily water withdrawal $\{w_i\}$ and of the external temperature $\{T_{e,i}\}$. Such sample profiles are defined following specific stochastic generation rules, described in Section II.C. No stochastic generation is adopted for the cold water temperature profile $\{T_o\}$, since its value is usually constant during the day.

The simulation of the $N$ sample EWHs yields $N$ profiles of the thermostat state $\{q_i^j\}$, $j = 1,2,...,N$, from which power consumption of the $i$-th aggregate $P_{sA,i}$ is computed as:

$$P_{sA,i} = \frac{P_{sA,i}^n}{N} \sum_{j=1}^{N} q_i^j. \quad (3)$$

As for any Monte Carlo simulation, the sample number $N$ should be large enough to make the result statistically consistent. In this paper, $N = 10^5$. Given the power consumption of each sub-aggregate $P_{sA,i}$, the one of the entire EWHs aggregate $P_A$ [W] is computed as:

$$P_A = P_{sA,1} + P_{sA,2} + \cdots + P_{sA,M}. \quad (4)$$

*B. Flexibility evaluation*

For a fixed set of the data in points a), b) and c), defined for each of the sub-aggregates that compose the entire EWHs aggregate, the power consumption profile $P_A$ will depend on the adopted set-point $T_{sp}$, in other words, $P_A = P_A(T_{sp})$.

The aggregate flexibility is computed considering three different operating conditions (OC):

- *base-OC*, in which the default temperature set-point $T_{sp}^b$ is adopted, i.e. the one typically defined by users )in this paper $T_{sp}^b = 65$ °C).
- *max-OC*, in which the temperature set-point is set in order to make the top threshold of the thermostat equal to the maximal temperature limit, i.e. $T_{sp}^{max} = T_{max} - \Delta$;
- *min-OC*, in which the temperature set-point is set in order to make the down threshold of the thermostat equal to the *user acceptable minimum temperature* $T_0^{min}$, i.e. $T_{sp}^{min} = T_0^{min} + \Delta$; $T_0^{min}$ is defined as *the minimum initial value required to keep the hot water temperature larger than the user desired temperature $T^*$, after any water withdrawal occurrence, with a 99% probability*.

The computation of the minimal user acceptable temperature $T_m^0$ is described in Section II.E.

Given the three OCs, three aggregate power consumption profiles can be computed: the baseline profile $P_A^b = P_A(T_{sp}^b)$, and the two ones associated to max-OC, i.e. $P_A^{max} = P_A(T_{sp}^{max})$, and to the min-OC, i.e. $P_A^{max} = P_A(T_{sp}^{max})$, respectively. All these profiles are achievable by the EWH aggregate. However, because of the different distances of the three set-points from the incoming cold water temperatures $T_{o,i}$ and the external temperatures $T_{e,i}$, it will result that $P_A^{min} \leq P_A^b \leq P_A^{max}$. Thus, the aggregate power consumption can be potentially increased or decreased from the baseline $P_A^b$ up to $P_A^{max}$ or to $P_A^{min}$, respectively.

Flexibilities $\Delta P_A^+(\Delta t, \tau)$ and $\Delta P_A^-(\Delta t, \tau)$ can be therefore computed as follows:

- $\Delta P_A^+(\Delta t, \tau)$ equal to the minimum value of the difference $P_A^{max} - P_A^b$ computed within the time interval $[\tau, \tau + \Delta t]$;
- $\Delta P_A^-(\Delta t, \tau)$ equal to the minimum value of the difference $P_A^b - P_A^{min}$ computed within the time interval $[\tau, \tau + \Delta t]$.

It is worth remarking that the objective of the paper is to evaluate the flexibility of EWHs aggregates in terms of maximal availability of variation of their energy consumption (power variation per time interval) and that it is beyond the scope of the paper to study how the variation can be actually realized. However, it is clear that the idea behind the proposed method is that the EWHs temperature set-points can be varied (*e.g.* via remote control of "smart" EWH devices carried out by the aggregator) to provide the regulation service required by the TSO.

*C. Stochastic generation of the external temeperature profile*

The external temperature $T_e$ is the temperature of the ambient around the boiler, which is inside the house. Therefore, $T_e$ is different from the outside ambient temperature $T_a$. Since the available data is the average daily profile of $T_a$ (point c) in Section II.A), following rules are adopted to compute the sample profiles of $T_e$:

1. for each sample $j$, a minimum temperature threshold $T_{e,min}^j$ is randomly generated according to a uniform distribution within interval [18,20] °C, and a maximum temperature threshold $T_{e,max}^j$ is randomly generated according to a uniform distribution within interval [26,24] °C;
2. if $T_{e,min}^j \leq \overline{T}_{a,i} \leq T_{e,max}^j$, then $T_{e,i}^j = \overline{T}_{a,i}$;
3. if $\overline{T}_{a,i} \leq T_{e,min}^j$, then $T_{e,i}^j = T_{e,min}^j$ (since it is supposed that all houses are equipped with air heating systems);
4. for a percentage $p_c$ [%] of samples, if $\overline{T}_{a,i} \geq T_{e,max}^j$, then $T_{e,i}^j = T_{e,max}^j$ (supposing that the $p_c$ % of houses are equipped with air cooling systems);
5. for the remaining percentage $100 - p_c$, if $\overline{T}_{a,i} \geq T_{e,max}^j$, then $T_{e,i}^j = \overline{T}_{a,i}$.

*D. Stochastic generation of the hot water withdrawal profile*

The average hot water withdrawal daily profile $\{\overline{w}\}$ is usually defined with a granularity of one hour by a percentage profile $\{\overline{w}^\%(h)\}$, $h = 0,1,...,23$, and the average value of daily hot water demand $\overline{w}^d$ [l] ($\overline{w}(h) = \overline{w}^d \overline{w}^\%(h)/100$).

To generate a sample profile of hot water withdrawal, it is assumed that a single withdrawal occurs for a time interval uniformly distributed between the values $\tau_{short} = 1$ min and $\tau_{long} = 10$ min, and with a flow uniformly distributed between the values $w_{min} = 4$ l/min and $w_{max} = 12$ l/min (corresponding to a quarter of the average flow of a tap and to the maximum average flow of a shower, respectively). Based on these assumptions, we can compute the average number of water withdrawals per hour as it follows:

$$\overline{n}(h) = \frac{\overline{w}^d \overline{w}^\%(h)}{100} \cdot \frac{4}{(\tau_{short} + \tau_{long})(w_{min} + w_{max})}. \quad (5)$$

The number of samples of the individual hot water withdrawals during hour $h$ is generated with a Poisson distribution with mean value $\overline{n}(h)$; moreover, the starting times of the individual water withdrawals are uniformly distributed within the hour with a granularity of one minute.

The flow of each single withdrawal is scaled up in order to represent the mix of hot and cold water operated by the users for obtaining the desired temperature $T^*$. Specifically, the flow $\widetilde{w}^j(\tau)$, computed for a given minute $\tau$, is scaled up as it follows:

$$w^j(\tau) = \widetilde{w}^j(\tau) \frac{T_o - T^*}{T_o - T_0^u} \quad (6)$$

where $T_0^u$ [°C] is the hot water temperature at the beginning of the single withdrawal.

*E. User acceptable minimum temperature computation*

To compute $T_0^{min}$, we need to first consider the hot water withdrawal by generating a statistical set of $L$ hot water withdrawal sample profiles with a time granularity of one minute $\{\widetilde{w}^j(\tau)\}$, $j = 1,...,L, \tau = 0,...,24 \cdot 60 - 1$, using the procedure described in Section II.D. The number $L$ should be sufficient high to give to the set of sample profiles a statistical consistency. In this study, $L = 5 \cdot 10^5$ has been used. Recall that $\widetilde{w}$ indicates the water flow obtained without applying the hot and cold water mix (6). Then, each of the sample profiles $\{\widetilde{w}^j(\tau)\}$ is processed as it follows:

$$\widetilde{W}^j(q) = 15 \cdot \sum_{k=0}^{14} \widetilde{w}^j(q \cdot 15 + k), \quad (7)$$

where $q = 0,...,95$. The set $\widetilde{W}^j(q)$ $j = 1,...,L$ constitutes a sample distribution of the water consumed in each quarter of hour $q$ of the day. We can therefore compute from these sample distributions the 99-th percentile, that we indicate with $\widetilde{W}_{99}(q)$, and apply the hot and cold water mix (6) to obtain:

$$W_{99}(q) = \widetilde{W}_{99}(q) \frac{T_o - T^*}{T_o - T_0^u}. \quad (8)$$

We know now that in the 99% of the cases the hot water consumption in the quarter of hour $q$ of the day will be smaller than $W_{99}(q)$. Under the hypothesis that: $T_0^u(q)$ is the initial temperature at the beginning of the quarter of hour $q$, $T_0^u(q)$ is equal to the lower threshold of the thermostat, and the thermal energy losses of the boiler shell and the EWH heating capacity are negligible during water withdrawals, from (1), we get:

$$\ln(T_f(q) - T_o) = \ln(T_0^u(q) - T_o) - \frac{\widetilde{W}_{99}(q)}{V} \frac{T^* - T_o}{T_0^u(q) - T_o}. \quad (9)$$

In (9) we have two monotonically increasing functions of the hot water temperature at the end of the quarter of hour $q$, $T_f(q)$, on the left, and of $T_0^u(q)$ on the right. Therefore, the minimum value of initial condition $T_0^u(q)$ to obtain that the hot water is larger than the desired temperature $T^*$ at the end of the quarter of hour $q$, namely $T_0^{min}(q)$, is obtained by setting $T_f(q) = T^*$ and numerically solving the following equation:

$$\ln(T^* - T_o) = \ln(T_0^{min}(q) - T_o) - \frac{\widetilde{W}_{99}(q)}{V} \frac{T^* - T_o}{T_0^{min}(q) - T_o}. \quad (10)$$

Finally, to assure that the condition holds true during all the day, a unique value of $T_0^{min}$ is computed as follows:

$$T_0^{min} = \max_{q=0,2,\dots,95} T_0^{min}(q). \quad (11)$$

## III. CASE STUDY

In this paper, potential aggregates of domestic EWHs in Italy are considered. According to the methodology introduced in Section II, we need to define the composition of the sub-aggregates, characterize them by defining the data in points a)-c), and identify the general data (equal for all sub-aggregates) in points d)-f).

### A. Sub-aggregates definintion

EWHs are aggregated into 15 aggregates, each covering a distinct geographical area.

### B. Types of EWHs

The population of HWHs in Italy can represented by three reference commercial models, whose technical data, sufficient to define parameters $V_i$, $R_i$, and $P_i^n$ (data in point a)), are reported in Table I (according to [13]). The table also reports the maximal temperature limit $T^{max}$ and the thermostat dead-band $2\Delta$ (data in points e) and f)).

### C. Sub-aggregates nominal powers

Table I reports the diffusion rate in Italy of the three classes of EWHs, estimated in [14]. Such distribution is assumed to be the same for all Italy. From the data available in [14], it is also possible to obtain the estimate of the number of HWHs installed in each aggregation area. This information is crossed with the EWHs data in Table I to obtain the nominal powers of the EWHs aggregates. In Table II we report the data of one of the aggregates, which corresponds to Sicily, and the ones of the entire Italian territory.

In Table II, there are some zeros, meaning that no district in a given area belongs to those climate zones. Therefore, in Table II, we see 9 sub-aggregates, related to nonzero entries. However, within each of these sub-aggregates, we supposed to have the three classes of EWHs, introduced in Section III.B, and distributed according to Table I. Thus, the total number of sub-aggregates are $M = 27$ and their nominal power $P_{sA,i}^n$ (data in point b)) can be computed by multiplying the aggregate total power reported in Table II, first with the climate zone percentage, always in Table II, and then with the EWH diffusion rate in Table I.

TABLE I. REFERENCE MODELS OF DOMESTIC EWHS IN ITALY

| Model | ARISTON PRO ECO R 50 V/3 | ARISTON PRO ECO R 80 V/3 | ARISTON PRO ECO R 100 V/3 |
|---|---|---|---|
| Capacity ($V$) | 50 l | 80 l | 100 l |
| Nominal Power ($P^n$) | 1.2 kW | 1.2 kW | 1.5 kW |
| Maximal temperature ($T^{max}$) | 75 °C | 75 °C | 75 °C |
| Thermal dispersion at 65°C | 0.99 kWh/d | 1.35 kWh/d | 1.56 kWh/d |
| Thermostat dead-band ($2\Delta$) | 5 °C | 5 °C | 5° C |
| Diffusion rate [14] | 22 % | 60 % | 18 % |

TABLE II. HWHS AGGREGATES NOMINAL POWERS

| Area | Installed power [MW] | Climate Zone [%] | | | | |
|---|---|---|---|---|---|---|
| | | B | C | D | E | F |
| **Sicily** | 934.76 | 85 | 6 | 5 | 3 | 0 |
| **Italy** | 4508.87 | 19.3 | 29.2 | 26.1 | 24.4 | 0.9 |

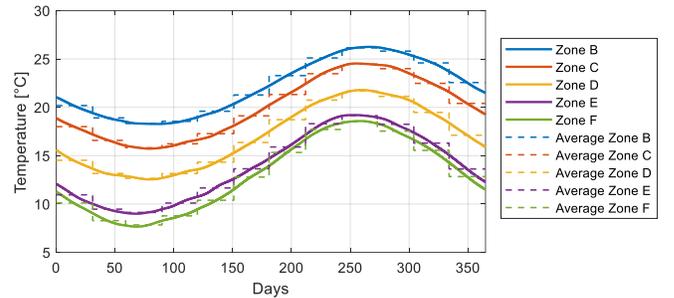

Figure 1. Cold water temperatures: one year profiles and monthly averages.

### D. Average daily profile of the outside ambient temperature

For each of the five climate zones and for each month of the year, an average daily profile of the external ambient temperature $\{\overline{T}_{a,i}\}$ has been computed, with time granularity of one hour (data in point c)). Such computation has been realized using the average temperatures registered in the five climate zones, for twenty years, collected in the database [15].

### E. Average daily profile of the cold water temperature

The average profiles of the cold water temperatures $\{\overline{T}_{o,i}\}$ (data in point c)) have been computed using the application TgCalc, developed by RSE and University of Padova [16]. The outputs of this application showed that there are negligible variations on the cold water temperature during a single day. Therefore, in order to characterize a typical day for each month, the monthly averages shown in Figure 1 have been computed and used to for all the five climate zones.

### F. Average daily profile of the hot water withdrawal

In the considered case study, domestic EHWs are considered. Therefore, the daily distribution profile of the hot water withdrawal $\{\overline{w}^\%(h)\}$ showed in Figure 2, is adopted for all the EHWs sub-aggregates. Such a profile is provided in [17] to represent the typical daily distribution of the domestic hot water demand. Also the value adopted for average total daily hot water demand $\overline{w}^d$ is the same for all the sub-aggregates and it is equal to 142 l. This value has been computed by crossing the information about the houses sizes in Italy (collected from [18]) and their correlation with the hot water demand, established in [19].

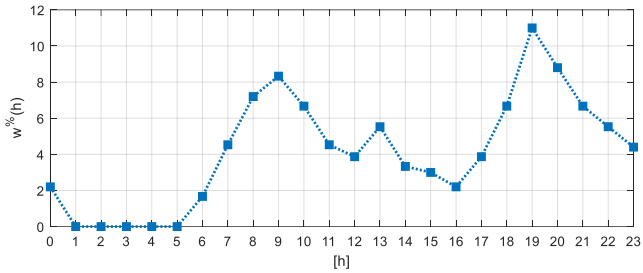

Figure 2. Hourly distribution of hot water withdrawals.

## IV. RESULTS

Figure 3 and Figure 4 show the positive and negative flexibilities $\Delta P_A^+(\Delta t, \tau)$ and $\Delta P_A^-(\Delta t, \tau)$ of Sicily, computed for four time extensions $\Delta t = 15, 30, 45, 60$ min, with starting times $\tau = q \cdot 15$ min, $q = 0, 1, ..., 95$, and for four months, one for each season. According to Table II, territory of Sicily mainly belongs to the climate zone B. It is the aggregation area with the largest total nominal power, equal to 934.76 MW.

In general, we observe that negative flexibility is larger than the positive ones. In both the cases, a specific shape of the flexibility profiles can be recognized. It is evidently related to the hot water withdrawal daily distribution (Figure 2).

Positive flexibility is larger during deep night (from 3 a.m. to 5 a.m.) and during the central daytime hours (around noon). This happens because, from 1 a.m., hot water demand is zero. Thus, after a couple of hours, most of EWHs are switched off and, consequently, they become ready to be activated. A similar scenario occurs around noon: after the morning peak of the hot water demand (9 a.m.), many EWHs recover the temperature set-point and switch off, becoming ready to be activated.

Negative flexibility is larger in the early morning (from 5 a.m. to 7 a.m.) and in the late afternoon (from 5 p.m. to 7 p.m.). This happens because from 5 a.m. the hot water demand starts increasing from zero to the morning peak at 9 a.m.. During the first two hours, many EWHs are switched on and become ready to be temporarily disactivated. Such an availability is lowered at the morning demand peak time, since, even if lots of EWHs are switched on, they cannot be easily deactivated because of the high requirement of hot water. A similar behavior occurs when, at 4 p.m., the hot water demand start increasing toward the evening peak at 7 p.m..

The maximal value of the positive flexibility is reached in August with about 22 MW in the night, whereas, in the other months, the night and daytime peaks are all at about 15 MW. In all the months, positive flexibility is zero during a time interval from 5 a.m. to 10 a.m.. Excepting for peak values, there are no significant differences among the positive flexibilities computed in the four months.

Maximal values of negative flexibility are reached in August with about 35 MW in the morning and about 27 MW in the afternoon. May and November are similar, with peaks of about 33 MW, whereas February is the month with the lower flexibility, especially in the late afternoon. Except for August, in all months, negative flexibility is zero during a time interval from 7 a.m. and 11 a.m..

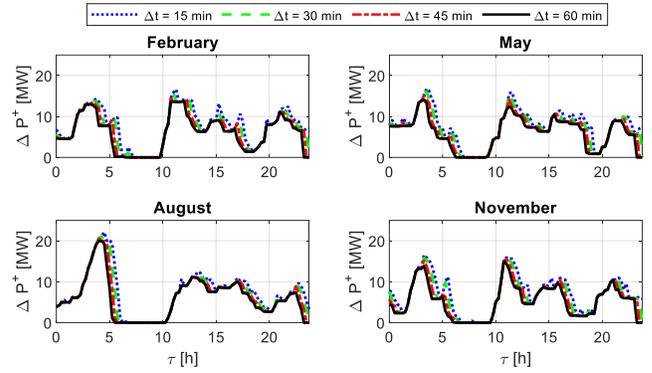

Figure 3. Positive flexibilities $\Delta P_A^+(\Delta t, \tau)$ of Sicily.

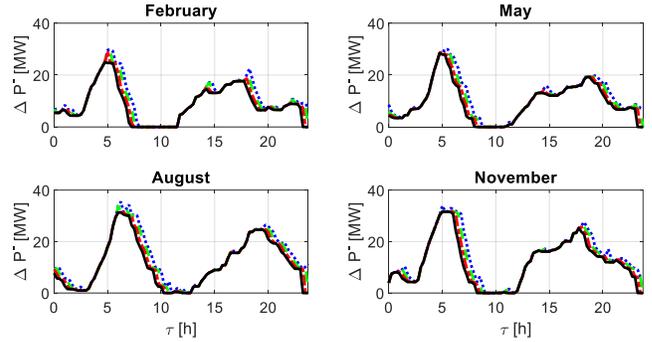

Figure 4. Negative flexibilities $\Delta P_A^-(\Delta t, \tau)$ of Sicily.

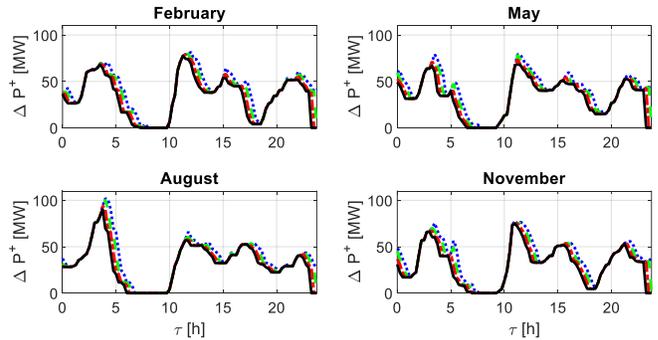

Figure 5. Positive flexibilities $\Delta P_A^+(\Delta t, \tau)$ of the overall Italy aggregate.

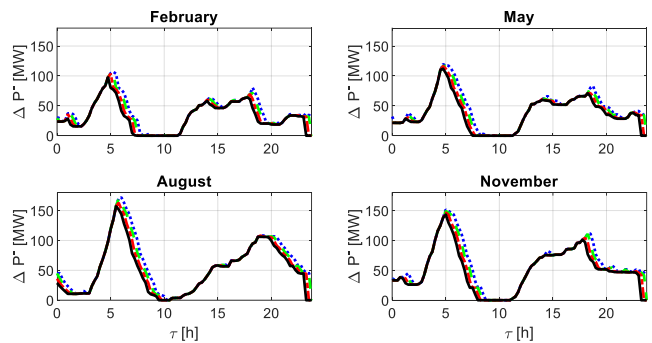

Figure 6. Negative flexibilities $\Delta P_A^-(\Delta t, \tau)$ of the overall Italy aggregate.

It is interesting to observe how flexibility depends on the time extension Δt. When the flexibility (both positive and negative) profiles are increasing, there is no dependence on the time extension; whereas, the dependence is more significant when profiles are decreasing. In this latter case, the largest values of flexibility are obtained with the minimum time extension of 15 min (dotted blue lines in the figures), whereas the smallest values are obtained with the maximum time extension of 60 min (solid black lines in the figures). In general, the difference is of the order of a few MW (up to 5 MW between 15 min to 60 min, registered in August for negative flexibility).

Figure 5 and Figure 6 show the positive and negative flexibilities of the aggregate composed by all the EWHs aggregates. According to Table II, the total nominal power is about 4.5 GW. As verified for Sicily, negative flexibility of the overall Italy aggregate is generally larger than positive flexibility. Both reach their peak in August, about 100 MW in the positive case, and about 170 MW in the negative case. The shapes of flexibility profiles are similar to the ones observed for Sicily, meaning that they mainly depends on the hot water withdrawal daily distribution (Figure 2), which is the same for all areas.

As for Sicily, except for August, the amount of the positive flexibility is similar in all the considered months, with night and daytime peaks of about 70 MW. The negative flexibilities of May and February are similar, with a morning peak of about 100-110 MW and a late afternoon peak of about 80 MW. Differently from Sicily, November flexibility is similar to the one of August, except for the peak value, which in November is lower, about 150 MW against the 170 MW of August.

The dependence of flexibility on the time extension $\Delta t$ is similar to the one verified for Sicily. The difference between the shortest time (15 min) and the longest one (60 min) is, in the overall Italy case, of the order of few MW (up to 10 MW between 15 min to 60 min, registered in August for negative flexibility).

## V. Conclusions

In this paper, a method to evaluate the flexibility of aggregates of domestic EWHs has been developed. Flexibility is defined as the potential capability of the aggregate to vary its power consumption from a baseline profile, for a given time extension. In particular, the results about one of the fifteen aggregates and of the overall Italy aggregate, in a typical day for each season, have been analyzed.

General conclusions are summarized in the following:

- positive flexibility (load increase) is generally lower than the negative one (load decrease);
- flexibility severely depends on time, because of the variation of the users demand of hot water during the day;
- negative flexibility depends on seasons, resulting to be generally higher in summer and lower in winter;
- positive flexibility does not strongly depend on seasons;
- flexibility slightly changes with the required time extension of the power variation, showing a difference of few MWs from the shortest considered time (15 min) to the longest one (60 min), also in the overall Italy aggregate case, with respect to power variations of the order of more than 100 MW.